\documentclass[prd,preprint,superscriptaddress,showpacs,byrevtex]{revtex4}
\usepackage{bm}
\def\fsl#1{\setbox0=\hbox{$#1$}           
   \dimen0=\wd0                                 
   \setbox1=\hbox{/} \dimen1=\wd1               
   \ifdim\dimen0>\dimen1                        
      \rlap{\hbox to \dimen0{\hfil/\hfil}}      
      #1                                        
   \else                                        
      \rlap{\hbox to \dimen1{\hfil$#1$\hfil}}   
      /                                         
   \fi}                                         %
\newcommand{\be}{\begin{equation}}
\newcommand{\ee}{\end{equation}}
\newcommand{\bea}{\begin{eqnarray}}
\newcommand{\eea}{\end{eqnarray}}
\newcommand{\beq}{\begin{equation}}
\newcommand{\eeq}{\end{equation}}
\newcommand{\beqs}{\begin{eqnarray}}
\newcommand{\eeqs}{\end{eqnarray}}

\newcommand{\aslash}{A\hspace{-0.067in}\slash}

\begin{document}
\title{ Correct Formulation Of Lattice QCD Method To Study Hadron Formation From Quarks and Gluons }
\author{Gouranga C Nayak }\thanks{E-Mail: nayakg138@gmail.com}
%
%
\date{\today}
\begin{abstract}
The present lattice QCD method can not study the hadron formation from the quarks and gluons. This is because it operates the unphysical QCD Hamiltonian of all the quarks plus antiquarks plus gluons inside the hadron on the physical energy eigenstate of the hadron to obtain the physical energy eigenvalue of the hadron which is not correct because of the non-vanishing boundary surface term in the energy conservation equation in QCD due to the confinement of quarks, antiquarks and gluons inside the finite size hadron. In this paper we present the correct formulation of the lattice QCD method to study the physical hadron formation from the unphysical quarks, antiquarks and gluons inside the hadron by using this non-vanishing boundary surface term in the lattice QCD.
\end{abstract}
\maketitle
\pagestyle{plain}

\pagenumbering{arabic}

\section{ Introduction }

A hadron (such as proton, neutron, pion) is a composite particle consisting of quarks, antiquarks and gluons which are the fundamental particles of the nature. The fundamental theory of the nature which describes the interaction between the quarks and the gluons is known as the quantum chromodynamics (QCD) which is the quantum field theory of the classical Yang-Mills theory \cite{yme}.

Unlike quantum electrodynamics (QED) where the photons do not directly interact with each other the gluons directly interact with each other in QCD. The gauge field lagrangian in QED contains quadratic powers of the photon field but the gauge field lagrangian in QCD contains quadratic, cubic and quartic powers of the gluon field. Because of the presence of the cubic and quartic powers of the gluon field in the QCD lagrangian it becomes impossible to solve the full QCD analytically.

In the renormalized QCD \cite{tve} the asymptotic freedom \cite{gwe,poe} predicts that the QCD coupling decreases as distance decreases and increases as the distance increases. Hence there has been extensive calculation in the short distance partonic scattering cross section at the high energy colliders by using the perturbative QCD (pQCD). By using the factorization theorem in QCD \cite{fce,fce1,fce2} this short distance partonic scattering cross section is convoluted with the (experimentally extracted) long distance parton distribution function inside the hadron and with the long distance parton to hadron fragmentation function to predict the hadronic cross section at the high energy colliders which is experimentally measured.

However, since the renormalized QCD coupling increases as the distance increases the pQCD is not applicable at the long distance where the non-perturbative QCD becomes applicable. Since the hadron formation from the quarks, antiquarks and gluons involves long distance physics one finds that the hadron formation from the quarks, antiquarks and gluons should be studied by using the non-perturbative QCD.

However, the non-perturbative QCD has not been solved analytically. This is because of the presence of the cubic and quartic powers of the gluon field in the QCD lagrangian which makes it impossible to perform the path integration in QCD analytically [see section \ref{latpath}]. But the path integration in QCD can be done numerically in the Euclidean time. Because of this reason the hadron formation from the quarks, antiquarks and gluons is studied by using the lattice QCD method by performing the path integration in QCD numerically in the Euclidean time.

One of the crucial equation which is used to study the hadron $H$ formation from the partons in the lattice QCD method is given by \cite{lte}
\bea
H^{\rm H}_{\rm Partons}|H_n(P)>= E_n^H|H_n(P)>
\label{eve}
\eea
where $H^{\rm H}_{\rm Partons}$ is the unphysical QCD Hamiltonian of all the partons inside the hadron $H$, the $|H_n(P)>$ is the physical energy eigenstate of the hadron $H$ in its n$th$ level normalized to unity, $E_n^H$ is the physical energy eigenvalue of the hadron $H$ in its n$th$ level and ${\vec P}$ is the physical momentum of the hadron $H$. The QCD Hamiltonian $H^{\rm H}_{\rm Partons}$ of all the partons inside the hadron $H$ is unphysical because we have not directly experimentally observed the quarks and gluons. The $|H_n(P)>$, $E_n^H$ and ${\vec P}$ of the hadron $H$ are physical because we have directly experimentally observed the hadrons.

Since the left hand side of eq. (\ref{eve}) is unphysical and the right hand side of eq. (\ref{eve}) is physical one finds that the eq. (\ref{eve}) must be wrong, {\it i. e.},
\bea
H^{\rm H}_{\rm Partons}|H_n(P)>\neq E_n^H|H_n(P)>.
\label{evne}
\eea
In fact eq. (\ref{eve}) is not consistent with the energy conservation equation in QCD obtained from the gauge invariant Noether's theorem in QCD which predicts that \cite{nkge}
\bea
\frac{d}{dt}<H(P)|H^{\rm H}_{\rm Partons}|H(P)>=\frac{dE^{\rm H}_{\rm Partons}(t)}{dt}=-\frac{dE_{\rm flux}(t)}{dt}\neq 0
\label{fve}
\eea
where $|H(P)>=|H_0(P)>$ is the physical energy eigenstate of the hadron $H$ in its ground state, $E^{\rm H}_{\rm Partons}(t)$ is the energy of all the quarks plus antiquarks plus gluons inside the hadron $H$ in its ground state and $E_{\rm flux}(t)$ is energy flux in QCD which is non-zero due to the confinement of quarks and gluons inside the finite size hadron $H$ \cite{nkfe, nkee}.

But in eq. (\ref{eve}) we have
\bea
\frac{dE^H}{dt}=0.
\label{gve}
\eea
Hence from eqs. (\ref{fve}) and (\ref{gve}) we find that the eq. (\ref{eve}) must be wrong. On the other hand the eqs. (\ref{fve}) and (\ref{gve}) are consistent with eq. (\ref{evne}). From eq. (\ref{fve}) we find
\bea
H^{\rm H}_{\rm Partons}|H_n(P)>=E^{\rm H}_{n,~{\rm Partons}}(t)|H_n(P)>
\label{jve}
\eea
which is consistent with eq. (\ref{evne}) where $E^{\rm H}_{n,~{\rm Partons}}(t)$ is the energy of all the quarks plus antiquarks plus gluons inside the hadron $H$ in its n$th$ level.

From eqs. (\ref{gve}) and (\ref{fve}) we find that the energy $E^H$ of the hadron $H$ is given by
\bea
E^H=E^{\rm H}_{\rm Partons}(t)+E_{\rm flux}(t)
\label{ive}
\eea
where $E^{\rm H}_{\rm Partons}(t)$ is the energy of all the quarks plus antiquarks plus gluons inside the hadron $H$ and $E_{\rm flux}(t)$ is energy flux in QCD which is non-zero due to the confinement of quarks and gluons inside the finite size hadron $H$ \cite{nkfe, nkee}. Hence we find that the energy of the hadron $H$ is not equal to the energy of all the quarks plus antiquarks plus gluons inside the hadron $H$ because of the presence of non-zero energy flux in QCD due to the confinement of quarks, antiquarks and gluons inside the finite size hadron $H$.

The present lattice QCD method in the literature operates the unphysical QCD Hamiltonian $H^{\rm H}_{\rm Partons}$ of all the quarks plus antiquarks plus gluons inside the hadron on the physical energy eigenstate $|H_n(P)>$ of the hadron to obtain the physical energy eigenvalue $E_n^H$ of the hadron as given by eq. (\ref{eve}) which is not correct. For example, for the pion $\pi^+$ formation from the quarks, antiquarks and gluons the present lattice QCD method uses eq. (\ref{eve}) to predict [see section \ref{latpath}]
\bea
&&m^2_{\pi^+}f^2_{\pi^+}~e^{-t m_{\pi^+}} = [<\Omega|\sum_{\vec x}{\cal O}_{\pi^+}({\vec x},t){\cal O}_{\pi^+}(0)|\Omega>]_{t \rightarrow \infty}
\label{pmei}
\eea
where $m_{\pi^+}$ is the mass of the $\pi^+$, the $f_{\pi^+}$ is the decay constant of $\pi^+$, the $|\Omega>$ is the non-perturbative QCD vacuum state and ${\cal O}_{\pi^+}(x)$ is the partonic operator for $\pi^+$ formation given by
\bea
{\cal O}_{\pi^+}(x) = d^\dagger_k(x) \gamma_5 u_k(x)
\label{pid}
\eea
where $u_k(x)$ and $d_k(x)$ is the Dirac field for the up and down quarks respectively with $k=1,2,3$ being the color index.

However, as discussed above, because of the non-vanishing boundary surface term [the non-zero energy flux $E_{\rm flux}(t)$ in eq. (\ref{fve})], the eq. (\ref{eve}) is not correct in QCD. Since eq. (\ref{eve}) is not the correct equation in QCD but eq. (\ref{jve}) is the correct equation in QCD one must use eq. (\ref{jve}) instead of eq. (\ref{eve}) to study the hadron formation from the quarks and gluons. In this paper we present the correct formulation of the lattice QCD method by using the correct equation (\ref{jve}) instead of the incorrect equation (\ref{eve}) to study the physical hadron formation from the unphysical quarks, antiquarks and gluons inside the hadron by using the non-zero energy flux $E_{\rm flux}(t)$ in the lattice QCD method.

We find that the correct formulation of the lattice QCD method which uses the correct equation (\ref{jve}) [instead of the incorrect equation (\ref{eve})] in QCD to study the pion $\pi^+$ formation from the quarks, antiquarks and gluons predicts
\bea
&&m^2_{\pi^+}f^2_{\pi^+}~ e^{-tm_{\pi^+}}=\left[\frac{<\Omega|\sum_{\vec x} {\cal O}_{\pi^+}({\vec x},t){\cal O}_{\pi^+}(0)|\Omega>}{e^{ [\frac{<\Omega|\sum_{{\vec x}'}~{\cal O}_{\pi^+}({\vec x}',t') [\int d^4x \sum_{q,{\bar q}, g} \partial_i T^{i0}_{\rm Partons}({\vec x},t)] {\cal O}_{\pi^+}(0)|\Omega>}{<\Omega|\sum_{{\vec x}'}{\cal O}_{\pi^+}({\vec x}',t') {\cal O}_{\pi^+}(0)|\Omega>}]_{t'\rightarrow \infty}}}\right]_{t \rightarrow \infty}
\label{pmcdi}
\eea
where the $\int dt$ is an indefinite integration and $\int d^3x$ is definite integration in $\int d^4x=\int dt \int d^3x$, the ${\cal O}_{\pi^+}(x)$ is given by eq. (\ref{pid}) and the $\sum_{q,{\bar q}, g}T^{i0}_{\rm Partons}(x)$ is the energy-momentum tensor density operator of all the quarks plus antiquarks plus gluons inside the pion $\pi^+$ with
\bea
&& T^{\nu \eta}_{\rm Partons}(x) =F^{\nu \lambda d}(x)F_\lambda^{~\eta d}(x) +\frac{1}{4} g^{\nu \eta} F^{\lambda \mu d}(x) F_{\lambda \mu}^d(x)+{\bar u}_l(x)\gamma^\nu [\delta^{lk} i\partial^\eta -igT^d_{lk}A^{\eta d}(x)]u_k(x)\nonumber \\
&& +{\bar d}_l(x)\gamma^\nu [\delta^{lk} i\partial^\eta -igT^d_{lk}A^{\eta d}(x)]d_k(x) +(antiquarks)
\label{tmdf}
\eea
and
\bea
F_{\lambda \nu}^h(x)=\partial_\lambda A_\nu^h(x) - \partial_\nu A_\lambda^h(x) +gf^{hds} A_\lambda^d(x) A_\nu^s(x)
\label{fme}
\eea
where $A_\mu^a(x)$ is the gluon field.

Hence we find that eq. (\ref{pmei}) is not the correct equation to study the $\pi^+$ formation from the quarks, antiquarks and gluons in the lattice QCD method but eq. (\ref{pmcdi}) is the correct equation to study the $\pi^+$ formation from the quarks, antiquarks and gluons in the lattice QCD method.

It is straightforward to extend eq. (\ref{pmcdi}) to other hadrons such as to proton, neutron and kion etc.

In this paper we will present a derivation of eq. (\ref{pmcdi}).

The paper is organized as follows. In section II we discuss the present lattice QCD method in the literature to study the hadron formation from the quarks and gluons and derive eq. (\ref{pmei}). In section III we discuss non-vanishing boundary surface term and the non-conservation of energy of partons inside the hadron due to the confinement in QCD. In section IV we present the correct formulation of the lattice QCD method to study the hadron formation from the quarks and gluons and derive eq. (\ref{pmcdi}). Section V contains conclusions.

\section{ Lattice QCD Method in the Literature To Study Hadron Formation From Quarks and Gluons }\label{latpath}

In this section we discuss the present lattice QCD method in the literature which uses the incorrect eq. (\ref{eve}) in QCD to study the physical hadron formation from the unphysical quarks and gluons. In section \ref{clatpath} we will formulate the correct lattice QCD method by using the correct eq. (\ref{jve}) instead of the incorrect eq. (\ref{eve}) to study the physical hadron formation from the unphysical quarks and gluons.

The vacuum expectation of the non-perturbative correlation function of the partonic operators ${\cal O}_a(x')...{\cal O}_b(x'')...{\cal O}_c(x''')$ in QCD is given by \cite{mte,abe}
\bea
&& <\Omega|{\cal O}_a(x')...{\cal O}_b(x'')...{\cal O}_c(x''')|\Omega>=\frac{1}{Z[0]} \int [d{\bar \psi}][d\psi] [dA] \times {\cal O}_a(x')...{\cal O}_b(x'')...{\cal O}_c(x''')\nonumber \\
&&\times {\rm det}[\frac{\delta B_f^d}{\delta \omega^h}] \times~{\rm exp}[i\int d^4x [-\frac{1}{4} F_{\lambda \nu}^h(x)F^{\lambda \nu h}(x) -\frac{1}{2\alpha} [B_f^h(x)]^2 \nonumber \\
&&+{\bar \psi}_k(x)[\delta^{kj}(i{\not \partial}-m)+gT^h_{kj}\aslash^h(x)]\psi_j(x)]]
\label{npe}
\eea
where $\psi_j(x)$ is the quark field, $A_\nu^h(x)$ is the gluon field, $B_f^h(x)$ is the gauge fixing term, $\alpha$ is the gauge fixing parameter, $m$ is the mass of the quark, $k.j=1,2,3$ is the color index of the quark field, $d,h=1,...,8$ is the color index of the gluon field, the partonic operator ${\cal O}_a(x)$ is a function of the quark and gluon fields, $|\Omega>$ is the non-perturbative QCD vacuum state which is different from the pQCD vacuum state $|0>$, the non-abelian gluon field tensor $F_{\lambda \nu}^h(x)$ is given by eq. (\ref{fme}) and the generating functional $Z[0]$ in the absence of the external sources is given by
\bea
&& Z[0]= \int [d{\bar \psi}][d\psi] [dA] \times {\cal O}_a(x')...{\cal O}_b(x'')...{\cal O}_c(x''')\times {\rm det}[\frac{\delta S_f^d}{\delta \omega^h}] \times {\rm exp}[i\int d^4x [\nonumber \\
&&-\frac{1}{4} F_{\lambda \nu}^h(x)F^{\lambda \nu h}(x) -\frac{1}{2\alpha} [B_f^h(x)]^2 +{\bar \psi}_k(x)[\delta^{kj}(i{\not \partial}-m)+gT^h_{kj}\aslash^h(x)]\psi_j(x)]].
\label{z0e}
\eea
In eqs. (\ref{npe}) and (\ref{z0e}) we do not have ghost fields because we directly work with the ghost determinant ${\rm det}[\frac{\delta B_f^d}{\delta \omega^h}]$ in this paper. A typical choice of the gauge fixing term is the covariant gauge fixing term $B_f^d(x)=\partial^\lambda A_\lambda^d(x)$ in the pQCD calculation at the high energy colliders.

Due to the presence of the cubic and quartic gluon field terms in the QCD lagrangian in eq. (\ref{npe}) it is not possible to perform the path integration in QCD analytically. But the path integration in QCD in eq. (\ref{npe}) can be performed numerically in the Euclidean time. Because of this reason the path integration in QCD in eq. (\ref{npe}) is performed numerically by using the lattice QCD method in the Euclidean time.

In this paper we consider the pion $\pi^+$ formation from the quarks, antiquarks and gluons using the lattice QCD method. The extension of this lattice QCD method to the other hadrons such as to the proton, neutron and kion etc. is straightforward.

The partonic operator ${\cal O}_{\pi^+}(x)$ for the pion $\pi^+$ formation is obtained from the up quark and the down antiquark fields as given by eq. (\ref{pid}) which has the same quantum number of the pion $\pi^+$. In order to study the pion $\pi^+$ formation from the partons we need to evaluate the vacuum expectation of the partonic two-point non-perturbative correlation function of the type $<\Omega|{\cal O}_{\pi^+}(x){\cal O}_{\pi^+}(0)|\Omega>$ which in the path integral formulation of the QCD is given by \cite{mte,abe}
\bea
&& <\Omega|{\cal O}_{\pi^+}(x'){\cal O}_{\pi^+}(0)|\Omega> =\frac{1}{Z[0]} \int [d{\bar u}][du][d{\bar d}[dd][dA] \times {\cal O}_{\pi^+}(x'){\cal O}_{\pi^+}(0) \times {\rm det}[\frac{\delta B_f^d}{\delta \omega^b}] \nonumber \\
&& \times ~ {\rm exp}[i\int d^4x [-\frac{1}{4} F_{\nu \eta}^b(x) F^{\nu \eta b}(x) -\frac{1}{2\alpha} [B_f^d(x)]^2+{\bar u}_k(x)[\delta^{kl}(i{\not \partial}-m_u)+gT^d_{kl}\aslash^d(x)]u_l(x)\nonumber \\
&&+{\bar d}_k(x)[\delta^{kl}(i{\not \partial}-m_d)+gT^d_{kl}\aslash^d(x)]d_l(x)]]
\label{cfd}
\eea
where $m_u$ is the mass of the up quark, $m_d$ is the mass of the down quark, the partonic operator ${\cal O}_{\pi^+}(x)$ for the pion $\pi^+$ formation is given by eq. (\ref{pid}) and the source free generating functional $Z[0]$ in QCD for the $\pi^+$ formation is given by
\bea
&& Z[0]= \int [d{\bar u}][du][d{\bar d}[dd][dA] \times {\rm det}[\frac{\delta B_f^d}{\delta \omega^b}] \times {\rm exp}[i\int d^4x [-\frac{1}{4} F_{\nu \eta}^b(x) F^{\nu \eta b}(x) -\frac{1}{2\alpha} [B_f^d(x)]^2 \nonumber \\
&&+{\bar u}_k(x)[\delta^{kl}(i{\not \partial}-m_u)+gT^d_{kl}\aslash^d(x)]u_l(x)+{\bar d}_k(x)[\delta^{kl}(i{\not \partial}-m_d)+gT^d_{kl}\aslash^d(x)]d_l(x)]].
\label{z0d}
\eea
For the pion $\pi^+$ case lattice QCD uses from the incorrect eq. (\ref{eve})
\bea
H^{\pi^+}_{\rm Partons}|\pi_n^+(P)>= E_n^{\pi^+}|\pi_n^+(P)>
\label{peve}
\eea
where ${\vec P}$ is the momentum of the pion $\pi^+$, the $H^{\pi^+}_{\rm Partons}$ is the QCD Hamiltonian of all the partons inside the pion $\pi^+$, the $|\pi_n^+(P)>$ is the energy eigenvector of the $\pi^+$ in its n$th$ level and $E_n^{\pi^+}$ is the energy eigenvalue of the $\pi^+$ in its n$th$ level.

Inserting complete set of pion energy eigenstates
\bea
\sum_n |\pi^+_n><\pi^+_n|=1
\label{csd}
\eea
and then using eq. (\ref{peve}) one finds from eq. (\ref{cfd}) in the Euclidean time
\bea
&& <\Omega|\sum_{\vec x}e^{-i{\vec P}\cdot {\vec x}} ~{\cal O}_{\pi^+}({\vec x},t){\cal O}_{\pi^+}(0)|\Omega> = \sum_n <\Omega|~{\cal O}_{\pi^+}(0)|\pi^+_n(P)><\pi^+_n(P)|{\cal O}_{\pi^+}(0)|\Omega>\nonumber \\
&&\times ~e^{-t E_n^{\pi^+}}.
\label{dse}
\eea

Assuming that all the higher level energy contributions in the large time limit $t \rightarrow \infty$ are negligible one finds that the ground state contribution dominates in eq. (\ref{dse}) to find
\bea
&& [<\Omega|\sum_{\vec x}e^{-i{\vec P}\cdot {\vec x}} ~{\cal O}_{\pi^+}({\vec x},t){\cal O}_{\pi^+}(0)|\Omega>]_{t \rightarrow \infty} = |<\Omega|~{\cal O}_{\pi^+}(0)|\pi^+(P)>|^2 ~e^{-t E^{\pi^+}}
\label{ete}
\eea
where $|\pi^+(P)>$ is the energy eigenstate of the pion $\pi^+$ in its ground state and $E^{\pi^+}$ is the energy of the pion $\pi^+$ in its ground state. For the pion $\pi^+$ at rest $({\vec P}=0)$ we find from eq. (\ref{ete})
\bea
&& [<\Omega|\sum_{\vec x}{\cal O}_{\pi^+}({\vec x},t){\cal O}_{\pi^+}(0)|\Omega>]_{t \rightarrow \infty} = |<\Omega|~{\cal O}_{\pi^+}(0)|\pi^+>|^2 ~e^{-t m_{\pi^+}}
\label{fte}
\eea
where $m_{\pi^+}$ is the mass of the pion $\pi^+$ at rest and $|\pi^+>$ is the energy eigenstate of the pion $\pi^+$ at rest in its ground state.

The decay constant $f_{\pi^+}$ of the pion $\pi^+$ is related to the amplitude $<\Omega|~{\cal O}_{\pi^+}(0)|\pi^+>$ via the equation
\bea
<\Omega|~{\cal O}_{\pi^+}(0)|\pi^+>=m_{\pi^+}f_{\pi^+}.
\label{pdce}
\eea
Using eq. (\ref{pdce}) in (\ref{fte}) we find
\bea
&&m^2_{\pi^+}f^2_{\pi^+}~e^{-t m_{\pi^+}} = [<\Omega|\sum_{\vec x}{\cal O}_{\pi^+}({\vec x},t){\cal O}_{\pi^+}(0)|\Omega>]_{t \rightarrow \infty}
\label{pme}
\eea
which reproduces eq. (\ref{pmei}).

Hence the mass and decay constant of the pion $\pi^+$ can be calculated from the first principle by using the lattice QCD from the eq. (\ref{pme}) from the rate of the exponential fall-off in time and from the amplitude. This technique can be extended to the other hadrons such as to the proton, neutron and kion etc.

This is the the present lattice QCD method which uses the incorrect eq. (\ref{eve}) [or the eq. (\ref{peve}) for the pion $\pi^+$ case] in QCD in the literature to study the physical hadron formation from the unphysical quarks and gluons. In section \ref{clatpath} we will formulate the correct lattice QCD method by using the correct eq. (\ref{jve}) instead of the incorrect eq. (\ref{eve}) to study the physical hadron formation from the unphysical quarks and gluons.

\section{Non-vanishing boundary surface term and the Non-conservation of Energy of All The Partons Inside The Hadron }

From the gauge invariant Noether's theorem in QCD we find the continuity equation \cite{nkge}
\bea
\partial_\lambda T^{\lambda \nu}(x) =0
\label{cee}
\eea
where $T^{\nu \eta}(x)$ is the energy-momentum tensor density operator of the partons in QCD. For $\nu=0$ we find from eq. (\ref{cee}) that the energy conservation equation of all the partons inside the pion $\pi^+$ at rest is given by [see eq. (\ref{fve})]
\bea
\frac{d[E^{\pi^+}_{\rm Partons}(t)+E_{\rm flux}(t)]}{dt}=0
\label{phve}
\eea
where [see eq. (\ref{fve})]
\bea
E^{\pi^+}_{\rm Partons}(t)=<\pi^+|\sum_{q,{\bar q},g}\int d^3x T^{00}_{\rm Partons}(t,{\vec x})|\pi^+>=<\pi^+|H^{\pi^+}_{\rm Partons}|\pi^+>
\label{pfve}
\eea
is the energy of all the partons inside the pion $\pi^+$ where $T^{\nu \eta}_{\rm Partons}(x)$ is the energy-momentum tensor density operator of the partons inside the pion $\pi^+$ as given by eq. (\ref{tmdf}) and [see eq. (\ref{cee})]
\bea
\frac{dE_{\rm flux}(t)}{dt}= <\pi^+|\sum_{q,{\bar q},g}\int d^3x~ \partial_j T^{j0}_{\rm Partons}(t,{\vec x})|\pi^+>\neq 0
\label{pfle}
\eea
is the time rate of the energy flux $E_{\rm flux}(t)$ in QCD which is non-zero due to the confinement of quarks and gluons inside the finite size pion $\pi^+$ \cite{nkfe, nkee}.

Note that there is also non-zero momentum flux ${\vec p}_{\rm flux}(t)$ in QCD which is non-zero due to the confinement of quarks and gluons inside the finite size pion $\pi^+$ \cite{nkme} which we do not require in this paper. Similarly there is also non-zero angular momentum flux ${\vec J}_{\rm flux}(t)$ in QCD which is non-zero due to the confinement of quarks and gluons inside the finite size pion $\pi^+$ \cite{nkje} which we do not require in this paper. We only require the non-zero energy flux $E_{\rm flux}(t)$ in this paper as given by eq. (\ref{pfle}) which determines the energy $E^H$ of the hadron from eq. (\ref{ive}).

\section{ Correct Formulation Of Lattice QCD Method To Study the Hadron Formation From Quarks and Gluons }\label{clatpath}

We saw in section I that eq. (\ref{eve}) is not the correct equation in QCD but eq. (\ref{jve}) is correct equation in QCD due to the confinement of quarks and gluons inside the finite size hadron. For the pion $\pi^+$ the eq. (\ref{jve}) [see eq. (\ref{pfve})] gives
\bea
H^{\pi^+}_{\rm Partons}|\pi^+_n(P)>=E^{\pi^+}_{n,~{\rm Partons}}(t)|\pi^+_n(P)>
\label{pjve}
\eea
where $E^{\pi^+}_{n,~{\rm Partons}}(t)$ is the energy of all the partons inside the pion $\pi^+$ in its n$th$ level. Using eqs. (\ref{pjve}) and (\ref{csd}) in (\ref{cfd}) we find in the Euclidean time
\bea
&& <\Omega|\sum_{\vec x}e^{-i{\vec P}\cdot {\vec x}} ~{\cal O}_{\pi^+}({\vec x},t){\cal O}_{\pi^+}(0)|\Omega> = \sum_n <\Omega|~{\cal O}_{\pi^+}(0)|\pi^+_n(P)><\pi^+_n(P)|{\cal O}_{\pi^+}(0)|\Omega>\nonumber \\
&&\times ~e^{-\int dt E^{\pi^+}_{n,~{\rm Partons}}(t)}
\label{dsd}
\eea
where $\int dt$ is an indefinite integration.

If all the higher energy level contribution exponentially falls off rapidly to zero at the larger time $t\rightarrow \infty$ then the ground state contribution dominates in which case we find from eq. (\ref{dsd})
\bea
&&[<\Omega|\sum_{\vec x}e^{-i{\vec P}\cdot {\vec x}} ~{\cal O}_{\pi^+}({\vec x},t){\cal O}_{\pi^+}(0)|\Omega>]_{t \rightarrow \infty}  =|<\Omega|~{\cal O}_{\pi^+}(0)|\pi^+(P)>|^2~e^{-\int dt E^{\pi^+}_{\rm Partons}(t)}\nonumber \\
\label{esd}
\eea
where $|\pi^+(P)>$ is the energy eigenstate of the pion $\pi^+$ in its ground state and $E^{\pi^+}_{\rm Partons}(t)$ is the energy of all the partons inside the pion $\pi^+$ in its ground state.

The energy $E^{\pi^+}$ of the pion $\pi^+$ and the energy $E^{\pi^+}_{\rm Partons}(t)$ of all the partons inside the pion $\pi^+$ are related by [see eq. (\ref{ive})]
\bea
E^{\pi^+}=E^{\pi^+}_{\rm Partons}(t)+E_{\rm flux}(t)
\label{efd}
\eea
where $E_{\rm flux}(t)$ is the energy flux [the boundary surface term] in QCD which is non-zero due to the confinement quarks, antiquarks and gluons inside the finite size pion $\pi^+$ \cite{nkfe, nkee}. Using eq. (\ref{efd}) in (\ref{esd}) we find for the pion $\pi^+$ at rest (${\vec P}=0$)
\bea
&&[<\Omega|\sum_{\vec x} {\cal O}_{\pi^+}({\vec x},t){\cal O}_{\pi^+}(0)|\Omega>]_{t \rightarrow \infty}  =|<\Omega|~{\cal O}_{\pi^+}(0)|\pi^+>|^2~e^{-tm_{\pi^+}}\times e^{\int dt E_{\rm flux}(t)}\nonumber \\
\label{pmd}
\eea
where $|\pi^+>$ is the energy eigenstate of the pion $\pi^+$ at rest, $m_{\pi^+}$ is the mass of the pion $\pi^+$ at rest and $E_{\rm flux}(t)$ is the energy flux [the boundary surface term] in QCD which is non-zero due to the confinement quarks, antiquarks and gluons inside the finite size pion $\pi^+$ at rest \cite{nkfe, nkee}.

The non-zero energy flux $E_{\rm flux}(t)$ in QCD due to the confinement quarks, antiquarks and gluons inside the finite size pion $\pi^+$ at rest is a non-perturbative quantity in QCD. Since the analytical solution of the non-perturbative QCD is not known the non-zero energy flux $E_{\rm flux}(t)$ in QCD can be calculated numerically by using the lattice QCD method in the Euclidean time.

In order to calculate $\frac{dE_{\rm flux}(t)}{dt}$ in eq. (\ref{pfle}) we need to calculate $<\pi^+|\sum_{q,{\bar q},g}\int d^3x \partial_j T^{j0}_{\rm Partons}(t,{\vec x})|\pi^+>$ in the lattice QCD method. For this purpose we need to evaluate the vacuum expectation value of the non-perturbative partonic three-point correlation function $<\Omega|\sum_{{\vec x}'}~{\cal O}_{\pi^+}({\vec x}',t') [\int d^3x \sum_{q,{\bar q}, g} \partial_i T^{i0}_{\rm Partons}({\vec x},t)] {\cal O}_{\pi^+}(0)|\Omega>$ and the vacuum expectation value of the non-perturbative partonic two-point correlation function $<\Omega|\sum_{{\vec x}'}{\cal O}_{\pi^+}({\vec x}',t') {\cal O}_{\pi^+}(0)|\Omega>$ in the lattice QCD method.

By using the path integral formulation of the lattice QCD method we find that the non-zero energy flux $E_{\rm flux}(t)$ in QCD due to the confinement quarks, antiquarks and gluons inside the finite size pion $\pi^+$ at rest is given by \cite{nkbse}
\bea
 \frac{dE_{\rm flux}(t)}{dt} = [\frac{<\Omega|\sum_{{\vec x}'}~{\cal O}_{\pi^+}({\vec x}',t') [\int d^3x \sum_{q,{\bar q}, g} \partial_i T^{i0}_{\rm Partons}({\vec x},t)] {\cal O}_{\pi^+}(0)|\Omega>}{<\Omega|\sum_{{\vec x}'}{\cal O}_{\pi^+}({\vec x}',t') {\cal O}_{\pi^+}(0)|\Omega>}]_{t'\rightarrow \infty}
\label{pefd}
\eea
where the partonic operator ${\cal O}_{\pi^+}(x)$ for the $\pi^+$ is given by eq. (\ref{pid}) and $T^{\mu \nu}_{\rm Partons}(x)$ is given by eq. (\ref{tmdf}).

Using eqs. (\ref{pefd}) and (\ref{pdce}) in (\ref{pmd}) we find
\bea
&&m^2_{\pi^+}f^2_{\pi^+}~ e^{-tm_{\pi^+}}=\left[\frac{<\Omega|\sum_{\vec x} {\cal O}_{\pi^+}({\vec x},t){\cal O}_{\pi^+}(0)|\Omega>}{e^{ [\frac{<\Omega|\sum_{{\vec x}'}~{\cal O}_{\pi^+}({\vec x}',t') [\int d^4x \sum_{q,{\bar q}, g} \partial_i T^{i0}_{\rm Partons}({\vec x},t)] {\cal O}_{\pi^+}(0)|\Omega>}{<\Omega|\sum_{{\vec x}'}{\cal O}_{\pi^+}({\vec x}',t') {\cal O}_{\pi^+}(0)|\Omega>}]_{t'\rightarrow \infty}}}\right]_{t \rightarrow \infty}
\label{pmcd}
\eea
which reproduces eq. (\ref{pmcdi}) where the $\int dt$ is an indefinite integration and $\int d^3x$ is definite integration in $\int d^4x=\int dt \int d^3x$. The eq. (\ref{pmcd}) can be extended to other hadrons such as to proton, neutron and kion etc.

Hence we find that eq. (\ref{pme}) which is used in the present lattice QCD method in the literature is not the correct equation to study the hadron formation from the quarks and gluons but eq. (\ref{pmcd}) which we have obtained in this paper is the correct equation to study the hadron formation from the quarks and gluons by in the lattice QCD method.

\section{Conclusions}

The present lattice QCD method can not study the hadron formation from the quarks and gluons. This is because it operates the unphysical QCD Hamiltonian of all the quarks plus antiquarks plus gluons inside the hadron on the physical energy eigenstate of the hadron to obtain the physical energy eigenvalue of the hadron which is not correct because of the non-vanishing boundary surface term in the energy conservation equation in QCD due to the confinement of quarks, antiquarks and gluons inside the finite size hadron. In this paper we have presented the correct formulation of the lattice QCD method to study the physical hadron formation from the unphysical quarks, antiquarks and gluons inside the hadron by using this non-vanishing boundary surface term in the lattice QCD.


\begin{thebibliography}{99}

\bibitem{yme} C. N. Yang and R. Mills, Phys. Rev. 96 (1954) 191.

\bibitem{tve} G. 't Hooft and M.J.G. Veltman, Nucl.Phys. B44 (1972) 189.

\bibitem{gwe} D. J. Gross and F. Wilczek, Phys. Rev. Lett. 30 (1973) 1343.

\bibitem{poe} D. Politzer, Phys. Rev. Lett. 30 (1973) 1346.

\bibitem{fce} J. C. Collins, D. E. Soper and G. Sterman, Nucl. Phys. B261 (1985) 104.

\bibitem{fce1} G. C. Nayak, J. Qiu and G. Sterman, Phys. Lett. B613 (2005) 45; Phys. Rev. D72 (2005) 114012; Phys. Rev. D74 (2006) 074007; Phys. Rev. D77 (2008) 034022.

\bibitem{fce2} G. C. Nayak, JHEP 1709 (2017) 090; Eur. Phys. J. C76 (2016) 448; Eur. Phys. J. Plus 133 (2018) 52; Phys. Part. Nucl. Lett. 13 (2016) 417; arXiv:1506.02651 [hep-ph]; Phys. Part. Nucl. Lett. 14 (2017) 18; J. Theor. Appl. Phys. 11 (2017) 275; arXiv:1705.07913 [hep-ph].

\bibitem{lte} S. Hashimoto, J. Lahio and S. R. Sharpe, {\it Lattice Quantum Chromodynamics}, Particle Data Group (2017).

\bibitem{nkge} G. C. Nayak, arXiv:1802.07825 [hep-ph].

\bibitem{nkfe} G. C. Nayak, arXiv:1807.09158.

\bibitem{nkee} G. C. Nayak, arXiv:1804.07211 [hep-ph].

\bibitem{mte} T. Muta, {\it Foundations of Quantum Chromodynamics}, World Scientific lecture notes in physics-Vol. 5.

\bibitem{abe} L. F. Abbott, Nucl. Phys. B185 (1981) 189.

\bibitem{nkme} G. C. Nayak, arXiv:1804.02712 [hep-ph].

\bibitem{nkje} G. C. Nayak, arXiv:1803.08371 [hep-ph].

\bibitem{nkbse} G. C. Nayak, {\it Lattice QCD Method To Study Non-Vanishing Boundary Surface Term in QCD}, submitted for publication.

\end{thebibliography}
\end{document}